\documentclass[12pt, centerh1]{article}
\usepackage{natbib}

\usepackage{booktabs}

\usepackage{lettrine}

\usepackage{enumitem}
\setlist[itemize]{noitemsep}

\usepackage{hyperref}

\usepackage{placeins,textgreek,colonequals}
\usepackage{amsmath,color}
\usepackage{amsfonts}
\usepackage{graphicx,colonequals}
\usepackage{caption}
\usepackage{subcaption}
\usepackage{pbox}
\usepackage[utf8]{inputenc}
\usepackage[english]{babel}
\usepackage{amsthm}
\usepackage{subcaption}
\usepackage{breqn}

\theoremstyle{definition}
\newtheorem{definition}{Definition}[section]

\newcommand{\mL}{\mathcal{L}}
\newcommand{\bthet}{{\boldsymbol{\theta}}}
\newcommand{\bTheta}{{{\pmb\Theta}}}

\newcommand{\bfx}{{\bf x}}

\newcommand{\vecx}{\mathbf{x}}

\newcommand{\vecX}{\mathbf{X}}

\newcommand{\vecz}{\mathbf{z}}

\newcommand{\varthet}{\mbox{\boldmath$\vartheta$}}
\newcommand{\varthe}{\mbox{\boldmath$\theta$}}

\title{Clustering Discrete-Valued Time Series}
\author{Tyler Roick$^*$, Dimitris Karlis$^{**}$ and Paul D. McNicholas$^*$}
\date{\small $^*$Department of Mathematics \& Statistics, McMaster University, Ontario, Canada.\\ $^{**}$Department of Statistics, Athens University of Economics and Business, Athens, Greece.}

\begin{document}
\maketitle{}

\begin{abstract}
There is a need for the development of models that are able to account  for discreteness in data, along with its time series properties and correlation.  Our focus falls on INteger-valued AutoRegressive (INAR) type models. The INAR type models can be used in conjunction with existing model-based clustering techniques to cluster discrete-valued time series data. With the use of a finite mixture model, several existing techniques such as the selection of the number of clusters, estimation using expectation-maximization and model selection are applicable. The proposed model is then demonstrated on real data to illustrate its clustering applications.\\[-10pt]

\end{abstract}

\section{Introduction}


Consider the case in which we have $n$ individuals observed at
certain time points. These data are then considered to be comprised
of $n$ time series, which is a typical panel data situation.
Suppose that we are interested in clustering the $n$ individuals
on a number of, say, $G$ clusters based on their observed time
series, i.e., based on the data  $\{ y_{it} \}$,
$i=1,\ldots,n,~t=1,\ldots,T_i$. Finally, consider that the data
are discrete, i.e., $y_{it} \in \{0,1,\ldots \}$, so that we
observed $n$ discrete-valued time series and we aim at clustering
the observations based on the characteristics of their series.

Such data may occur in certain circumstances. For example,
consider the situation where individuals count their respective
number of drinks per day for a certain time period aiming at
identifying different drinking patterns among individuals. The
goal is to cluster individuals based on their different drinking
patterns. This example will be discussed in depth later. In
accident analysis, the goal is to cluster sites with similar
accident history in order to develop before and after studies
which measure the effect of an intervention. In consumer research,
the goal is to use the time series of different consumers and
their daily/weekly purchases of a product in order to cluster them
based on purchasing patterns. It should be emphasized that since
the observation can be a small number of counts, the discreteness
of the data needs to be handled with care.

Clustering time series has been a problem that has attracted much
research, especially for the case of time series that take
continuous values. In this paper, the aim is to establish and
apply model-based clustering to appropriately defined discrete
valued time series. Model-based clustering  for time series has
been applied in the past, see the work of
\citet[][]{fruhwirth2008model} and the references therein.
%
As opposed to distance-based clustering methods, model-based
clustering using finite mixture models extends to time series in a
quite natural way. Model-based clustering of time series may be
based on many different classes of finite mixture models.

Model-based clustering is a technique for estimating group memberships,
 in which no observations are {\it a priori} labeled, based on parametric
 finite mixture models. Finite mixture models are based on the assumption that a population
  is a convex linear combination of a finite number of densities. A random vector $\vecX$
  is said to arise from a parametric finite mixture distribution if, for all $\vecx \subset \vecX$, its density can be written as
  $$f(\vecx \ | \ \varthet) = \sum_{g=1}^{G} \pi_g f_g(\vecx \ |\ \varthe_g),$$ where $\pi_g > 0$, such that $\sum_{g=1}^{G}\pi_g = 1$,
  is called the $g$th mixing proportion, $f_g(\vecx \ |\ \varthe_g)$ is the $g$th component density, and $\varthet = (\pi_1, \ldots, \pi_{G-1}, \varthe_1, \ldots, \varthe_G)$ is the vector of parameters. The component densities $f_1(\vecx \ |\ \varthe_1), f_2(\vecx \ |\ \varthe_2), \ldots, f_G(\vecx \ |\ \varthe_G)$ are usually taken to be of the same type. 
In our case, due to the discreteness of the data, they will be
joint probability mass functions that reflect the underlying time
series model. Further details and an in-depth review of
model-based clustering can be found in
\citet[][]{mcnicholas16a,mcnicholas16b}.


In this paper, a novel model-based approach for clustering discrete-valued time series is introduced.
The approach utilizes a finite mixture of INteger AutoRegressive (hereafter INAR) processes to cluster the data. 
The work done in the current literature \citep{schnatter10a,
schnatter10b,schnatter11} involves model-based clustering of
categorical time series based on time-homogeneous first-order
Markov chains, model-based clustering of panel or longitudinal
data based on finite mixture models, and model-based clustering of
categorical time series with multinomial logit classification.

 Note that the literature contains certain other
approaches for clustering time series. Some such approaches are
based on a distance between the series \citep{d2019trimmed} and others on
some other characteristics of the series, e.g., in the
periodogram \citep{caiado2006periodogram} or the autocorrealtion
function \citep{d2009autocorrelation}. There are also approaches that use a
combination of the aforementioned methods such as the approach in
\cite{alonso2019clustering}, where some kind of distance is
calculated based on the autocorrelation. Finally, there is work on model-based
methods as, for example, in \cite{xiong2004time} by using ARMA
mixture models and as already mentioned in
\citet{fruhwirth2008model}. The interested reader can consult very
nice reviews by \cite{aghabozorgi2015time} and
\cite{caiado2015time}, while a more detailed treatment of the
problem can be found in \cite{maharaj2019time} and the references
therein. We emphasize that our paper focuses on the discrete
nature of the data as it is captured by a model-based approach.
Existing methods not created for discrete data need to be
explained in more detail when the data are counts, as in our case.
Later, in Section~$3$, we make use of such a method that proposes
fuzzy clustering \citep{Izakian15} mainly for illustration and
comparison purposes.

In Section~2, the framework of our methodology for clustering
discrete-valued time series via a mixture of INAR models is
presented. The implementation of the EM algorithm for parameter
estimation, convergence, initialization, model selection, and
performance assessment will be covered.
In Section~3, our methodology is applied to both simulated and
real data sets, and the results of the application are discussed.
In Section~4, a discussion of the work presented throughout this
paper is given. Thoughts on the direction of future work are also
considered.

\section{Methodology}

\subsection{INAR($p$) model}

In recent times, there has been an increasing number of
applications for discrete-valued time series models
\citep[see][]{weiss2018introduction}. The models developed herein
are based on a class of INAR models. Certain other models can
also be considered for each group.

\theoremstyle{definition}
\begin{definition}{(INAR$(1)$ Process with Binomial Thinning)}\label{def1}
A discrete time non-negative integer-valued process
$\{X_t\}_{\mathbb{Z}}$ is said to be a INAR$(1)$ process if it
satisfies the  recursion $$X_{t} = \alpha \circ
X_{t-1}+\epsilon_{t}, $$ where $\alpha \in [0,1]$, the symbol
$\circ$ represents the binomial thinning operator, and
$\{\epsilon_t\}_{\mathbb{Z}}$ is a sequence of non-negative
i.i.d.\ integer-valued random variables with mean $\mu_{\epsilon}$
and variance $\sigma_{\epsilon}^{2}$. All thinning operations are
performed independently of each other and of
$\{\epsilon_t\}_{\mathbb{Z}}$, and the thinning operations at each
time $t$ and $\epsilon_{t}$ are independent of $\{X_s\}_{s<t}$.
\end{definition}

\theoremstyle{definition}
\begin{definition}{(Binomial Thinning)}\label{def2}
Let $X$ be a non-negative integer-valued random variable and let
$\alpha \in [0,1]$. Define the random variable $$\alpha \circ X
\colonequals \sum_{i=1}^{X} Y_i,$$ where the $Y_i$ are i.i.d.\
Bernoulli indicators according to B(1, $\alpha$), which are also
independent of $X$. It can then be said that $\alpha \circ X$
arises from $X$ by binomial thinning.
\end{definition}

Binomial thinning was introduced by \citet[][]{steutel79} to
accommodate ``self-decomposability" and  ``stability" for
integer-valued time series. This becomes important in regards to
the INAR$(1)$ process, which will be discussed following thinning
operators. Recall that $\circ$ represents the binomial thinning
operator defined in Definition~\ref{def2}. Additionally, let
$\mu_X = \mathbb{E}[X]$ and $\sigma_X^{2} =
\mathbb{V}\text{ar}[X]$, then some basic properties of binomial
thinning, with proofs provided by \citet[][]{freeland98} and
\citet[][]{dasilva05a}, are as follows:
\begin{equation*}\begin{split}
   &\mathbb{E}[\alpha \circ X] = \alpha \mu_X,\\
   &\mathbb{V}\text{ar}[\alpha \circ X] = \alpha^{2} \sigma_X^{2} + \alpha(1-\alpha) \mu_X,\\
   &\text{Cov}[\alpha \circ X, X] =\alpha \sigma_X^{2}.
\end{split}\end{equation*}
See \cite{weiss08} for other generalizations of the thinning
operator.

The above definition of the INAR$(1)$ model depends on the
distribution of the innovation term $\epsilon_{t}$. The
distributional assumptions of $\epsilon_{t}$ result in the
marginal properties of the process. For example, assuming a
Poisson distribution for $\epsilon_{t}$ we end up with a time
series with Poisson marginals, which is perhaps not appropriate to
describe data with overdispersion (variance greater than the
mean). The model can be extended to what is referred to as the
INAR$(p)$ model.

\theoremstyle{definition}
\begin{definition}{(INAR$(p)$ Process)}\label{def3}
A discrete time non-negative integer-valued process
$\{X_t\}_{\mathbb{Z}}$ is said to be a INAR$(p)$ process if it
satisfies the following recursion \begin{dmath} $$ X_t =
\sum_{i=1}^{p} \alpha_i \circ X_{t-i}+\epsilon_t, $$ \end{dmath}
where $ \alpha_i \geq 0$ for $ i=1, \ldots, p-1$ and $\alpha_p >
0$.
\end{definition}

Further details on extensions of the model can be found in
\citet[][]{weiss2018introduction}. Note that the INAR$(p)$ model
can be interpreted in two different ways which may cause some
confusion. The above specification may either imply applying
binomial thinning sequentially and independently or applying a
multinomial type of thinning. These two approaches lead to
different models in many aspects (e.g., different marginals,
different likelihoods).

Finally, we write INAR$(p*)$ to denote the model $$X_t =  \alpha_p
\circ X_{t-p}+\epsilon_t.$$ In this model, lags up to and including
lag $p-1$ are excluded and only lag $p$ is considered. The reason
for such a model is that we can fit periodic autocorrelations. For
example, if the only autocorrelation in daily data comes at order
seven, then an INAR$(7*)$ model is a parsimonious one. This can
also be thought of as a INAR$(p)$ model with
$\alpha_1=\cdots=\alpha_{p-1}=0$.

\subsection{The Model}\label{sec:themodel}

Consider the INAR$(1)$ model defined in Definition~\ref{def1}. The
conditional likelihood of such a model, being a stationary Markov
chain,  can be written as
\begin{equation} \label{eq:3}
\mL(\bTheta) = P(X_1 ) \prod_{t=2}^{T}P(X_t | x_{t-1},\bTheta),
\end{equation}
where $\bTheta = (\alpha,\bthet)$ is the vector of parameters, $\alpha$ refers to the probability of success for binomial
thinning, and $\bthet = (\lambda, \phi)$ are the parameters
associated with the distribution of the  innovation terms. The
parameters $\lambda$ and $\phi$ refer to the mean and dispersion
of the innovations, respectively. Note that, for $t=1$, the term
$P(X_1)$ in \eqref{eq:3} refers only to the innovation
distribution.  Considering the previously given definition of
binomial thinning, the conditional distribution of the model can
be seen as a convolution between the binomial distribution and
that of the distribution of the innovation terms. The conditional
likelihoods for INAR($p$) processes are similar to that of
\eqref{eq:3}. The general conditional likelihood where the
observations are related at higher-order lag times, assuming the
same structure as the INAR$(1)$ process, can be written as
\begin{equation} \label{eq:4}
\mL(\bTheta) = \prod_{t=1}^{s}P(X_t)\prod_{t=s+1}^{T}P(X_t |
x_{t-s},\bTheta),
\end{equation}
where the first product of (\ref{eq:4}) corresponds to the
distribution of the innovation terms only. The likelihood in
(\ref{eq:4}) is the likelihood contribution for one individual
only.  For panel data, the product of all the individual
likelihoods must be introduced.

In the model-based clustering scenario, we assume a finite mixture
of  such likelihoods. Although the observations are assumed to
have come from an INAR process, they may come from any finite
mixture of INAR processes with equal or different orders. If the
INAR processes are of different orders, then the respective
clusters are considered to differ with respect to temporal
patterns. The observations are said to have come from a mixture of
INAR processes included in the model with a specific probability.
That is to say that each individual belongs to a specific INAR
process which does not change over time, but the process may have
different orders. The finite mixture of likelihoods for the INAR
model can be written as
\begin{equation} \label{eq:5}
\mL_i(\bTheta) = \sum_{g=1}^{G} \pi_g \mL_{ig}(\bTheta_g),
\end{equation}
where $\pi_g > 0$, such that $\sum_{g=1}^{G} \pi_g =1$, are the
mixing proportions, $\mL_i(\bTheta)$ refers to the likelihood for
the $i$th individual, $\mL_{ig}(\bTheta_g)$ refers to the
likelihood of the $i$th individual coming from the $g$th process,
and $\bTheta_g$ denotes the parameters of the $g$th INAR process
which can be of any order.

The likelihood for each individual is found over time from $t=1$
to $T_i$.  The number of components $G$ is considered to be
unknown and will be estimated using a criterion (see
Section~\ref{sec:thebic}). The finite mixture of likelihoods in
(\ref{eq:5}) can then be seen to follow a similar structure to the
definition of a mixture model given in Section~\ref{sec:themodel}.
An important note here is that the model assumes that each
individual has a certain INAR process and differs from the model
in \cite{bockenholt1998mixed}, where only the innovation term
follows a mixture of distributions. Being a finite mixture model
we can estimate it using the expectation-maximization (EM) algorithm  as described in the next
subsection.

\subsection{EM Algorithm}
The EM algorithm \citep{dempster77}
is an iterative procedure used to find maximum likelihood estimates
in the case of missing or incomplete data. Each iteration of the EM
algorithm involves two steps, the
expectation (E) step and the maximization (M) step. The E-step
involves computing the conditional expected value of the complete-data
log-likelihood, while the M-step maximizes the expected value of
the complete-data log-likelihood with respect to the model
parameters. Complete-data refers to the combination of the
observed and unobserved data. The E- and M-steps are iterated until
convergence is reached --- in practice, this means some stopping rule is satisfied (see below).

In straightforward applications of model-based clustering, the complete-data is comprised of the observed data $\vecx_1, \ldots, \vecx_n$ along with the unknown group membership labels $\vecz_1, \ldots, \vecz_n$, where $\vecz_i = (z_{i1}, \ldots, z_{iG})$ and 
\[
z_{ig}=
\begin{cases}
1 & \text{if} \ \bfx_i \ \text{belongs to component } g,\\
0 & \text{otherwise,} \\
\end{cases}
\]
for $i=1,\ldots,n$ and $g=1,\ldots,G$. The estimation of $z_{ig}$,
or specifically of $P(Z_{ig}=1)$, is the primary objective of
model-based clustering.

A well-known approach for determining whether the EM algorithm has
converged is by the  use of Aitken's acceleration
\citep{aitken26}. The Aitken acceleration procedure estimates the
asymptotic maximum log-likelihood at each iteration of the EM
algorithm and makes a decision about whether it has converged or
not. At iteration $k$, the Aitken acceleration is given by
\begin{equation*}
a^{(k)} = \frac{\ell^{(k+1)} - \ell^{(k)}}{\ell^{(k)} -
\ell^{(k-1)}},
\end{equation*}
where $\ell^{(k)}$ is the log-likelihood value from iteration $k$.
An asymptotic (i.e., after many iterations) estimate of the
log-likelihood at iteration $k+1$ is given by
\begin{equation*}
\ell_{\infty}^{(k+1)} = \ell^{(k)} + \frac{1}{1 - a^{(k)}}
(\ell^{(k+1)} - \ell^{(k)}).
\end{equation*}
The stopping criterion proposed by \citet{lindsay95} suggests that
the EM algorithm has converged when
\begin{equation} \label{eq:7}
\ell_{\infty}^{(k+1)} - \ell^{(k+1)} < \epsilon,
\end{equation}
where $\epsilon$ is a small value. An alternative stopping
criterion was proposed by \cite{mcnicholas10b}, which suggests
that the algorithm has converged when
\begin{equation} \label{eq:8}
\ell_{\infty}^{(k+1)} - \ell^{(k)} \in (0,\epsilon),
\end{equation}
for a small value of tolerance $\epsilon$. 
For fixed $\epsilon$, the criterion in \eqref{eq:8} is at least as strict as
\eqref{eq:7}. Further, the
criterion in \eqref{eq:8} is at least as strict as the lack of
progress criterion, i.e.,
$\ell^{(k+1)} - \ell^{(k)} < \epsilon$,
in the neighborhood of a maximum \citep{mcnicholas10b}.

\subsection{Model Fitting}

Considering that the model follows a similar structure to that of
the definition of a finite mixture model, estimation via the EM
algorithm is considered. As the focus of this method is for
model-based clustering purposes, the scenario in which there are
$n$ observations, none of which have known group memberships, is
also considered.
At each E-step, the component indicator variables are  updated
using their conditional expected values
\begin{equation} \label{eq:6}
\hat{z}_{ig} = \frac{\pi_g \mL_{ig}(\bTheta_g)}{\sum_{g=1}^{G}
\pi_g \mL_{ig}(\bTheta_g)} = \frac{\pi_g
\mL_{ig}(\bTheta_g)}{\mL_i(\bTheta)}.
\end{equation}
In each M-step, the (conditional) expected value of the complete-data log-likelihood is
maximized with respect to the model parameters. The mixing
proportions are first updated via
\begin{equation*}
\hat{\pi}_g = \frac{n_g}{n},
\end{equation*}
for $g=1,\ldots,G$, where $n_g = \sum_{i=1}^{n} \hat{z}_{ig}$. Because we do not have a closed form expression, the weighted likelihood
\begin{equation*}
\mL^{g}(\bTheta) = \sum_{i=1}^{n} z_{ig} \mL_{ig}(\bTheta_g),
\end{equation*}
can be maximized via the {\tt optim} function in {\sf R}
\citep{R18} to obtain the model-specific parameters. At each successive iteration of the above steps, the likelihood is increased until a set convergence condition is met.
The stopping criterion in \eqref{eq:8} is used to determine whether the EM algorithm has converged.

\subsection{Initialization}

For each number of components $G$, there must be initial values
for the $\bTheta_g$. The objective is to obtain the true values of
the model parameters to optimize $\hat{z}_{ig}$. The ability to
obtain good starting values for the parameters proves to be
heavily dependent on the distribution of the innovations. In the
case of equal dispersion, herein referred to as equidispersion,
the innovations are assumed to follow a Poisson distribution.

Equidispersion implies that $\phi=1$ in $$\mathbb{E}[X_i] = \phi
\mathbb{V}\text{ar}[X_i].$$ In the case of overdispersion ($\phi >
1$), the innovations can be thought to follow a negative binomial
distribution. Overdispersion is the result of the variance being
larger than the expectation.
It is worth noting that the weighted likelihood $\mL^{g}(\bTheta)$
frequently fails to be optimized if dispersion is not accounted
for and Poisson innovations are used. Although very rare, the case
of under-dispersion is handled similarly.

In all cases, starting values are obtained with the use  of $k$-means clustering. The initial values of the means $\lambda_g$ are taken to be similar to the first group of centers found by $k$-means. The mixing proportions $\pi_g$ are based on the respective cluster sizes. For $\phi=1$, the probability of success, $\alpha_g$, for the binomial distribution is estimated by minimizing the average of the absolute difference of sums between simulated data and that of the observed data for the clusters found by $k$-means. This is done using the previously estimated values of $\lambda_g$ and $\pi_g$, respectively. The simulated data that the observed data is compared to are created using the most influential lag time. A similar approach is used in the case of $\phi \neq 1$, although both $\phi_g$ and $\alpha_g$ must be estimated here. Minimizing the absolute mean of the difference between the observed data and simulated data provides moderately accurate
 starting values for both.  The model proves to be more accurate when initialized in an iterative fashion, meaning that initialization is really only an issue for the smallest number of components $G$ fitted. Subsequent numbers of components $G$ use the maximized parameter values found when $G-1$ components are fitted with the addition of a new component centered at the mean with a small mixing proportion. 

\subsection{Model Selection and Performance Assessment}\label{sec:thebic}

The models for this method are considered to be the possible
mixtures of  INAR processes. The INAR processes to be included in
the mixtures are decided by their respective autocorrelations. For
example, in Figure \ref{fig:TLFBBox}, the two most influential
autocorrelations are of order one and order seven, respectively. 
If these were
the only two desired autocorrelations to be included in the model,
then any mixture of these two autocorrelations may be used. This
means that the possible models are mixtures of the form $(G-H)$
INAR$(5*)$ and $H$ INAR$(10*)$, where $G$ is the number of
components and $H \leq G$. It is obvious that $H$ is restricted by
$G$ because a negative number of INAR processes cannot be fitted, but
as $G$ increases so does the total number of possible mixtures.

With the use of mixture models, an  objective criterion is needed
to select the `best' model.
The Bayesian information criterion \citep[BIC;][]{schwarz78} can
be used to select the best model.
Given a model with parameters $\bTheta$, the Bayesian information
criterion can be written
\begin{equation*}
\text{BIC} = 2 \ell(\hat{\bTheta}) - \rho\log n,
\end{equation*}
where $\ell(\hat{\bTheta})$ is the maximized log-likelihood,
$\hat{\bTheta}$ is the maximum likelihood estimate of $\bTheta$,
$\rho$ is the number of free parameters, and $n$ is the number of
observations.
For our model, in the case of Poisson distributed innovations,
there are $G$ free parameters from the estimation of $\lambda$,
$G$ from the estimation of $\alpha$, and $G-1$ from the estimation
of $\pi_g$. Note that, in the case of negative binomial distributed innovations, there are an additional $G$ free parameters from the
estimation of the dispersion $\phi$. The number of observations
comes from the total amount of time points for all individuals.

Although in a real clustering scenario the true group memberships
are not  known, the effectiveness of the model will be evaluated
through simulated data with known group memberships. The model is
evaluated using a cross-tabulation of the maximum {\it a
posteriori} (MAP) classification of the predicted group
memberships versus the true group memberships. Using the results
of the cross-tabulation, the performance can be quantified
numerically though the use of the adjusted Rand index
\citep[ARI;][]{hubert85}.
The ARI corrects the Rand index \citep{rand71} for agreement by
chance and has an expected value of $0$ under random
classification and a value of $1$ for perfect class agreement.



\section{Applications}

\subsection{Overview}
The model developed in Section~$2$ will now be applied in both
simulation studies and real data analyses. An analysis of the results
for each application will be carried out. The two simulation
studies reflect the different aspects covered throughout
Section~$2$ in regards to equidispersion and overdispersion. For
simplicity in the analyses, only the two most influential INAR
processes will be considered in the models. The INAR processes to
be included in the model will be decided by the most influential
autocorrelations at a multitude of different lag times. We will
also only consider two possible models in each analysis, i.e.,
mixtures of $(G-H)$ INAR$(i*)$ and $H$ INAR$(j*)$, where $i$ and
$j$ are the most influential autocorrelations, $G$ is the number
of components, and $H \in \{0,1\}$ . Due to two INAR processes and two
models being considered, $G=1$ components will not be fitted. This
is done for consistency purposes while following the iterative
approach mentioned previously and also corresponds to the kind of
models used later in the real data analysis.

For each simulated data analysis, $100$ simulated data sets will
be generated for each of five increasing difficulty scenarios.
To increase the challenge in clustering, the parameters of the
simulated data will converge together (as the difficulty increases) to bring the clusters closer
and create more overlap. Both simulated data analyses will be done
in a clustering fashion such that the true group memberships of
the data will be taken as unknown. This allows us to assess the
performance and classification accuracy using the ARI.

\subsection{Simulated Data Analyses}

\subsubsection{Poisson Innovation Simulated Data}\label{sec:poissoninn}

INAR data with Poisson distributed innovations are  simulated with
increasing difficulty. The true parameters along with the mixing
proportions of the two components in each simulation can be found
in Table~\ref{tab:PoissonClust}. In this case, $10,000$
two-component observations are simulated. The two-component model
in this scenario is simulated from a mixture of two INAR$(5*)$.
The dimensions of the simulated data are for $200$ individuals
over times $t=1,\ldots,50$.

Given that the data was simulated from a mixture of two
INAR$(5*)$, it  is to be expected that autocorrelations at lag $5$
and lag $10$ would be the most influential autocorrelations.
Hence, mixtures of INAR$(5*)$ and INAR$(10*)$ are used for the
model. Because the data have been simulated for Poisson
distributed innovations, the dispersion of the data will mainly
follow along the Poisson line, where $\phi=1$.

For each of five clustering scenarios with different overlapping
clusters, $G\in\{2,3\}$ components are fit using $k$-means starting
values.  The tolerance level for convergence is taken to be
$10^{-1}$. The results of each trial can be seen in Table
\ref{tab:PoissonClust} along with corresponding MAP
classifications. The BIC in almost all clustering difficulties
selects $G=2$ components using a mixture of two INAR$(5*)$ and
zero INAR$(10*)$ as the best model. This is prominent from the
collapsed classification tables that can be seen in Table
\ref{tab:PoissonClust}. The mean estimated parameters over the
$100$ simulations in each clustering difficulty are remarkably
close to the true parameters with the exception of the dispersion
in the very easy, easy, and moderate difficulties. The mean
dispersion is very skewed in these three difficulties as it only
appears larger than $\phi = 1$ in $8$, $5$, and $7$ cases,
respectively. The is due to the fact that some Poisson simulated
data sets are slightly overdispersed but fitted as if equidispersion were present. The overall difference
between true and estimated parameters is very marginal. It can be
seen that when a three-component model is selected, it is common 
that a very small mixing proportion is assigned to that component with parameters that are sensible in light of the first two components.  Note, the averages of the third component 
parameters are reported as an average over the number of cases in which a third 
component was selected.  The mean and standard deviation of the 
ARI for each of the $100$ simulations are reported in
Table~\ref{tab:PoissonClust}. All scenarios achieve very good
results for their respective difficulties. It can be seen from the
very difficult clustering scenario that results dwindle slightly.
Although, reasonable parameters are found and an overall
misclassification rate of $0.117$ is still achieved 
(Table~\ref{tab:PoissonClust}). The
classification table presents the true classification as rows and
those derived as columns accumulated over the 100 replications (200
observations for each replication, in total 20000 observations
classified for the whole experiment). 
\begin{table*}[!ht]
    \caption{Clustering results for the very easy, easy, moderate, difficult, and very difficult simulated INAR data with Poisson distributed innovations.}
    \label{tab:PoissonClust}
    \centering
    \begin{tabular}{ p{2cm} p{3.5cm} p{4.5cm} p{1cm} p{4cm}}
        \hline
        Clustering Difficulty & True \hspace{3cm} Parameters & Mean \hspace{3cm} Estimated Parameters & Mean ARI (SD) & Classification Table\\
        \hline
        Very Easy  & \vspace{-0.8cm} $(\alpha_1, \pi_1, \lambda_1, \phi_1) = (0.20,0.375,7.00,1)$ $(\alpha_2, \pi_2, \lambda_2, \phi_2) = (0.70,0.625,0.50,1)$ & \vspace{-0.8cm}$(\hat{\alpha}_1, \hat{\pi}_1, \hat{\lambda}_1, \hat{\phi}_1) = (0.201,0.375,7.05,1)$ $(\hat{\alpha}_2, \hat{\pi}_2, \hat{\lambda}_2, \hat{\phi}_2) = (0.708,0.625,0.52,3345)$ & $1$\hspace{3cm}$(0)$ &
        \begin{tabular}{c| c c}
            & 1 & 2  \\
            \hline
            1 &  7500 & 0 \\
            2 &   0 & 12500\\
        \end{tabular} \\
         \hline
        Easy  & \vspace{-0.8cm} $(\alpha_1, \pi_1, \lambda_1, \phi_1) = (0.40,0.375,6.00,1)$ $(\alpha_2, \pi_2, \lambda_2, \phi_2) = (0.70,0.625,0.50,1)$ & \vspace{-0.8cm}$(\hat{\alpha}_1, \hat{\pi}_1, \hat{\lambda}_1, \hat{\phi}_1) = (0.399,0.375,6.04,1)$ $(\hat{\alpha}_2, \hat{\pi}_2, \hat{\lambda}_2, \hat{\phi}_2) = (0.705,0.625,0.53,216)$ & $1$\hspace{3cm}$(0)$ &
        \begin{tabular}{c| c c}
            & 1 & 2  \\
            \hline
            1 &  7500 & 0 \\
            2 &   0 & 12500 \\
        \end{tabular} \\
        \hline
        Moderate & \vspace{-0.8cm} $(\alpha_1, \pi_1, \lambda_1, \phi_1) = (0.40,0.375,6.00,1)$ $(\alpha_2, \pi_2, \lambda_2, \phi_2) = (0.50,0.625,2.00,1)$ & \vspace{-0.8cm}$(\hat{\alpha}_1, \hat{\pi}_1, \hat{\lambda}_1, \hat{\phi}_1) = (0.405,0.371,6.01,1)$ $(\hat{\alpha}_2, \hat{\pi}_2, \hat{\lambda}_2, \hat{\phi}_2) = (0.506,0.625,2.01,3955)$ $(\hat{\alpha}_3, \hat{\pi}_3, \hat{\lambda}_3, \hat{\phi}_3) = (0.440,0.034,4.37,1)$ & $0.997$ \hspace{3cm} $(0.02)$ &
        \begin{tabular}{c| c c c}
            & 1 & 2 & 3 \\
            \hline
            1 &  7427 &  0 &  73 \\
            2 &  0 &  12500 &  0 \\
        \end{tabular} \\
        \hline
        Difficult & \vspace{-0.8cm} $(\alpha_1, \pi_1, \lambda_1, \phi_1) = (0.45,0.375,4.00,1)$ $(\alpha_2, \pi_2, \lambda_2, \phi_2) = (0.50,0.625,2.00,1)$ & \vspace{-0.8cm}$(\hat{\alpha}_1, \hat{\pi}_1, \hat{\lambda}_1, \hat{\phi}_1) = (0.452,0.373,4.03,1)$ $(\hat{\alpha}_2, \hat{\pi}_2, \hat{\lambda}_2, \hat{\phi}_2) = (0.503,0.625,2.02,1)$ $(\hat{\alpha}_3, \hat{\pi}_3, \hat{\lambda}_3, \hat{\phi}_3) = (0.421,0.026,4.49,1)$ & $0.997$ \hspace{3cm} $(0.01)$ &
        \begin{tabular}{c| c c c}
            & 1 & 2 & 3 \\
            \hline
            1 &  7479 &  8 &  13 \\
            2 &  1 &  12499 &  0 \\
        \end{tabular} \\
        \hline
        Very \newline Difficult & \vspace{-0.8cm} $(\alpha_1, \pi_1, \lambda_1, \phi_1) = (0.45,0.375,4.00,1)$ $(\alpha_2, \pi_2, \lambda_2, \phi_2) = (0.50,0.625,3.00,1)$ & \vspace{-0.8cm}$(\hat{\alpha}_1, \hat{\pi}_1, \hat{\lambda}_1, \hat{\phi}_1) = (0.467,0.370,3.91,1)$ $(\hat{\alpha}_2, \hat{\pi}_2, \hat{\lambda}_2, \hat{\phi}_2) = (0.483,0.627,3.10,1)$ $(\hat{\alpha}_3, \hat{\pi}_3, \hat{\lambda}_3, \hat{\phi}_3) = (0.53,0.028,3.76,1)$ & $0.594$ \hspace{3cm} $(0.12)$ &
        \begin{tabular}{c| c c c}
            & 1 & 2 & 3 \\
            \hline
            1 &  6261 &  1208 &  31 \\
            2 &  1106 &  11394 &  0 \\
        \end{tabular} \\
        \hline
    \end{tabular}
\end{table*}

\subsubsection{Negative Binomial Innovation Simulated Data}

Following in a similar fashion to Section~\ref{sec:poissoninn}, INAR data
with negative binomial distributed innovations are simulated with
increasing difficulty. The true parameters along with the mixing
proportions of the two components in each simulation can be found
in Table \ref{tab:NBClust}. In this case, $7,500$ two-component
observations are simulated. The two-component model in this
scenario is simulated from a mixture of one INAR$(2*)$ and one
INAR$(4*)$. The dimensions of the simulated data are for $250$
individuals over times $t=1,\ldots,30$.

Given that the data were simulated from the aforementioned
mixture, it is to be expected that autocorrelations at lag $2$ and
lag $4$ would be the most influential autocorrelations. Hence,
mixtures of INAR$(2*)$ and INAR$(4*)$ are used for the model.
Because the data have been simulated with negative binomial
distributed innovations, the dispersion of the data always lies
above the Poisson line, thus strongly indicating overdispersion.

For each of five clustering scenarios $G\in\{2,3\}$ components are fitted
using $k$-means starting values. The  tolerance level for
convergence is again set to $10^{-1}$. The results of each trial
can be seen in Table \ref{tab:NBClust} along with corresponding
MAP classifications. The BIC commonly selects $G=2$ components
using a mixture of one INAR$(2*)$ and one INAR$(4*)$ as the best
model. This model is selected in all $100$ simulations for the
very difficult clustering scenario. For the very easy and easy
clustering scenarios, it was seen that good performance could be
achieved using a mixture of two INAR$(2*)$ and zero INAR$(4*)$.
From the collapsed classification tables in Table~\ref{tab:NBClust}, the BIC has correctly selected the number of components in almost all simulations. Unlike the Poisson
distributed innovations, an ARI of over $0.94$ was achieved for all
clustering difficulties (Table~\ref{tab:NBClust}). This is quite
remarkable considering that existing clustering methods for time
series data are unable to account for overdispersion --- see, e.g., Section~\ref{sec:dtwfuzzy}. The mean estimated
parameters for the simulated data with negative binomial
distributed innovations appear to be slightly more accurate than
the Poisson distributed innovations. There also exists no skewed
results as the model is fit for data with overdispersion which is
always present in this scenario.
\begin{table*}[h!]
    \caption{Clustering results for the very easy, easy, moderate, difficult, and very difficult simulated INAR data with negative binomial distributed innovations.}
    \label{tab:NBClust}
    \centering
    \begin{tabular}{ p{2cm} p{3.5cm} p{4.5cm} p{1cm} p{4cm}}
        \hline
        Clustering Difficulty & True \hspace{3cm} Parameters & Mean \hspace{3cm} Estimated Parameters & Mean ARI (SD) & Classification Table\\
        \hline
        Very Easy  & \vspace{-0.8cm} $(\alpha_1, \pi_1, \lambda_1, \phi_1) = (0.80,0.60,1.00,4)$ $(\alpha_2, \pi_2, \lambda_2, \phi_2) = (0.20,0.40,9.00,2)$ & \vspace{-0.8cm}$(\hat{\alpha}_1, \hat{\pi}_1, \hat{\lambda}_1, \hat{\phi}_1) = (0.800,0.600,1.00,4.12)$ $(\hat{\alpha}_2, \hat{\pi}_2, \hat{\lambda}_2, \hat{\phi}_2) = (0.102,0.398,9.91,2.40)$ $(\hat{\alpha}_3, \hat{\pi}_3, \hat{\lambda}_3, \hat{\phi}_3) = (0.166,0.171,9.77,3.00)$ & $0.998$\hspace{3cm}$(0.02)$ &
        \begin{tabular}{c| c c c}
            & 1 & 2 & 3 \\
            \hline
            1 &  15000 & 0 & 0\\
            2 &   0 & 9957 & 43\\
        \end{tabular} \\
        \hline
        Easy  & \vspace{-0.8cm} $(\alpha_1, \pi_1, \lambda_1, \phi_1) = (0.70,0.60,3.00,4)$ $(\alpha_2, \pi_2, \lambda_2, \phi_2) = (0.20,0.40,9.00,2)$ &\vspace{-0.8cm}$(\hat{\alpha}_1, \hat{\pi}_1, \hat{\lambda}_1, \hat{\phi}_1) = (0.699,0.600,3.00,4.04)$ $(\hat{\alpha}_2, \hat{\pi}_2, \hat{\lambda}_2, \hat{\phi}_2) = (0.134,0.398,9.61,2.27)$ $(\hat{\alpha}_3, \hat{\pi}_3, \hat{\lambda}_3, \hat{\phi}_3) = (0.071,0.014,9.54,3.63)$ &  $0.997$\hspace{3cm}$(0.01)$ &
        \begin{tabular}{c| c c c}
            & 1 & 2 & 3 \\
            \hline
            1 &  15000 & 0 & 0\\
            2 &   0 & 9963 & 37\\
        \end{tabular} \\
        \hline
        Moderate & \vspace{-0.8cm} $(\alpha_1, \pi_1, \lambda_1, \phi_1) = (0.70,0.60,3.00,4)$ $(\alpha_2, \pi_2, \lambda_2, \phi_2) = (0.35,0.40,7.00,2)$  & \vspace{-0.8cm}$(\hat{\alpha}_1, \hat{\pi}_1, \hat{\lambda}_1, \hat{\phi}_1) = (0.699,0.600,3.00,4.07)$ $(\hat{\alpha}_2, \hat{\pi}_2, \hat{\lambda}_2, \hat{\phi}_2) = (0.353,0.398,6.98,1.99)$ $(\hat{\alpha}_3, \hat{\pi}_3, \hat{\lambda}_3, \hat{\phi}_3) = (0.101,0.005,9.33,7.04)$ & $0.993$ \hspace{3cm} $(0.01)$ &
        \begin{tabular}{c| c c c}
            & 1 & 2 & 3 \\
            \hline
            1 &  14988 &  8 &  4 \\
            2 & 16 &  9945 &  39 \\
        \end{tabular} \\
        \hline
        Difficult & \vspace{-0.8cm} $(\alpha_1, \pi_1, \lambda_1, \phi_1) = (0.50,0.60,4.00,4)$ $(\alpha_2, \pi_2, \lambda_2, \phi_2) = (0.35,0.40,7.00,2)$ & \vspace{-0.8cm}$(\hat{\alpha}_1, \hat{\pi}_1, \hat{\lambda}_1, \hat{\phi}_1) = (0.498,0.600,4.01,4.01)$ $(\hat{\alpha}_2, \hat{\pi}_2, \hat{\lambda}_2, \hat{\phi}_2) = (0.351,0.399,7.00,2.01)$ $(\hat{\alpha}_3, \hat{\pi}_3, \hat{\lambda}_3, \hat{\phi}_3) = (0.083,0.002,10.25,6.90)$ & $0.961$ \hspace{3cm} $(0.02)$ &
        \begin{tabular}{c| c c c}
            & 1 & 2 & 3 \\
            \hline
            1 &  14928 &  70 &  2 \\
            2 &  167 &  9821 &  12 \\
        \end{tabular} \\
        \hline
        Very \newline Difficult & \vspace{-0.8cm} $(\alpha_1, \pi_1, \lambda_1, \phi_1) = (0.50,0.60,4.00,4)$ $(\alpha_2, \pi_2, \lambda_2, \phi_2) = (0.40,0.40,6.00,2)$ & \vspace{-0.8cm}$(\hat{\alpha}_1, \hat{\pi}_1, \hat{\lambda}_1, \hat{\phi}_1) = (0.501,0.600,3.99,4.01)$ $(\hat{\alpha}_2, \hat{\pi}_2, \hat{\lambda}_2, \hat{\phi}_2) = (0.400,0.400,6.00,2.00)$ & $0.940$ \hspace{3cm} $(0.03)$ &
        \begin{tabular}{c| c c}
            & 1 & 2 \\
            \hline
            1 &  14884 &  116 \\
            2 &  267 &  9733 \\
        \end{tabular} \\
        \hline
    \end{tabular}
\end{table*}

\subsubsection{Result Comparison to Fuzzy Clustering}\label{sec:dtwfuzzy}

To provide a sensible comparison to the clustering approach
developed in Section~$2$, we have chosen to compare the
simulation study results to that of Fuzzy C-Medoids
\citep[FCMdd;][]{Krishnapuram01} clustering using dynamic time
warping distance \citep[DTW;][]{Berndt94}. Details of this
clustering approach for time series are described in
\citet{Izakian15}. In applying this approach, $G\in\{2,3\}$ 
components are fit using $5$ random starts and a fuzziness
parameter of $2$.  The tolerance level for convergence is set to a
more strict value than that of the INAR approach, i.e., it is taken to
be $10^{-2}$. This approach is applied to each of the $100$
simulated data sets for each of five clustering difficulties in
both the equidispersion and overdisperion scenarios. To summarize results, the average ARI
values, and the associated standard deviations, are reported alongside collapsed
classification tables.

In Table~\ref{tab:poissonfuzzy}, the results of the  FCMdd
approach for Poisson distributed innovations can be seen. The
FCMdd approach performs worse than the INAR approach for all $5$
clustering difficulties with a much larger variation in the
results. This approach frequently selects the incorrect number of
components in numerous simulations for each difficulty. The ability of the FCMdd approach to produce reasonably good results for the first four difficulties can most likely be attributed to the lack of overdispersion present in the data. 
\begin{table*}[!ht]
    \caption{FCMdd comparison for Poisson distributed innovations.}
    \label{tab:poissonfuzzy}
    \centering
    \begin{tabular}{ p{2cm}  p{5cm}  p{2.5cm}  p{2.5cm}}
        \hline
        Clustering Difficulty & FCMdd \hspace{3cm} Classification & Mean Fuzzy ARI (SD) &  Mean INAR ARI (SD) \\
        \hline
        Very Easy  &
        {\small
            \begin{tabular}{c| c c c}
                & 1 & 2 & 3 \\
                \hline
                1 &  7467 & 2 & 31 \\
                2 &  203  & 11476 & 821 \\
        \end{tabular} }  &
        $0.913 (0.18)$
        &  $1 (0)$ \\
        \hline
        Easy  &
        {\small
            \begin{tabular}{c| c c c}
                & 1 & 2 & 3 \\
                \hline
                1 &  7194 & 210 & 96 \\
                2 &   260 & 11573 & 667 \\
        \end{tabular} }  &
        $0.889(0.25)$
        &  $1 (0)$ \\
        \hline
        Moderate  &
        {\small
            \begin{tabular}{c| c c c}
                & 1 & 2 & 3 \\
                \hline
                1 &  7417 & 24 & 59 \\
                2 &   303 & 10317 & 1880  \\
        \end{tabular} }  &
        $0.820(0.22)$
        &  $0.997 (0.02)$ \\
        \hline
        Difficult  &
        {\small
            \begin{tabular}{c| c c c c}
                & 1 & 2 & 3\\
                \hline
                1 &  7084 & 245 & 171 \\
                2 &  368 & 11481 & 651 \\
        \end{tabular} }  &
        $0.844(0.22)$
        &  $0.997 (0.01)$ \\
        \hline
        Very \hspace{2cm} Difficult  &
        {\small
            \begin{tabular}{c| c c c}
                & 1 & 2 & 3 \\
                \hline
                1 &  5181 & 2100 & 219 \\
                2 &   3596 & 8592 & 312 \\
        \end{tabular} }  &
        $0.189(0.13)$
        &  $0.594 (0.12)$ \\
        \hline
    \end{tabular}
\end{table*}

In Table~\ref{tab:nbfuzzy}, the results of the FCMdd approach for negative binomial distributed innovations are presented. It can be seen that the INAR approach outperforms FCMdd for all five clustering difficulties to a greater extent than for the Poisson distributed innovations. The FCMdd approach is unable to account for overdispersion in the data. Other time series properties, and the possibility of large amounts of zeros in the data, may also be unaccounted for. Overall, considering that factors such as overdispersion are commonly seen in real data, the INAR model can be considered far superior when it comes to clustering discrete-valued time series.
\begin{table*}[!ht]
    \caption{FCMdd comparison for negative binomial distributed innovations.}
    \label{tab:nbfuzzy}
    \centering
    \begin{tabular}{ p{2cm}  p{5cm}  p{3cm}  p{2cm}}
        \hline
        Clustering Difficulty & FCMdd \hspace{3cm} Classification & Mean Fuzzy ARI (SD) &  INAR ARI \\
        \hline
        Very Easy  &
        {\small
            \begin{tabular}{c| c c c}
                & 1 & 2 & 3 \\
                \hline
                1 &  14937 & 5 & 58 \\
                2 &   132 & 9695 & 173 \\
        \end{tabular} }  &
        $0.968(0.05)$
        &  $0.998(0.02)$ \\
        \hline
        Easy  &
        {\small
            \begin{tabular}{c| c c c}
                & 1 & 2 & 3\\
                \hline
                1 & 14479 & 395 & 126 \\
                2 & 1178 & 8470 & 352 \\
        \end{tabular} }  &
        $0.750(0.12)$
        &  $0.997(0.01)$ \\
        \hline
        Moderate  &
        {\small
            \begin{tabular}{c| c c c}
                & 1 & 2 & 3 \\
                \hline
                1 &  13399 & 1297 & 304 \\
                2 &   1626 & 7975 & 399 \\
        \end{tabular} }  &
        $0.554(0.12)$
        &  $0.993(0.01)$ \\
        \hline
        Difficult  &
        {\small
            \begin{tabular}{c| c c c}
                & 1 & 2 & 3 \\
                \hline
                1 &  14104 & 546 & 350 \\
                2 &  2067 & 7378 & 555 \\
        \end{tabular} }  &
        $0.582(0.12)$
        &  $0.961(0.02)$ \\
        \hline
        Very \hspace{2cm} Difficult  &
        {\small
            \begin{tabular}{c| c c c}
                & 1 & 2 & 3 \\
                \hline
                1 &  12824 & 1988 & 188 \\
                2 &  2305 & 7395 & 300 \\
        \end{tabular} }  &
        $0.423(0.13)$
        &  $0.940(0.03)$ \\
        \hline
    \end{tabular}
\end{table*}

\subsection{Real Data Analyses}

\subsubsection{Alcohol Timeline Followback Data}

The timeline followback (TLFB) method \citep{sobell86} is a tool used to assess subjects' daily alcohol consumption. The alcohol TLFB data we consider \citep{atkins13} is available at {\tt www.researchgate.net},
and comes from a larger study aimed at event-specific prevention. The event-specific prevention here refers to intensive daily drinking habits around a number of people's twenty-first birthdays. This data also includes extreme drinking events relative to a random sample of students' drinking \citep{neighbors10}. Estimates of daily drinking were evaluated for clinical and nonclinical populations; e.g., adolescents, adults, college students, alcoholics of different severity, and normal male and female drinkers in the general population. Using a calendar, subjects provided retrospective estimates of their daily drinking over a specified time period. The original focus of the assessment was to study the gender, whether the subject is in a fraternity/sorority or neither (``greek status''), and period of the week in which the drinking occurred. Our focus will be solely on the number of drinks and what can be inferred about the clusters found.

\begin{figure}[!ht]
    \centering
    \begin{subfigure}{.45\linewidth}
        \centering
        \includegraphics[width=0.75\linewidth]{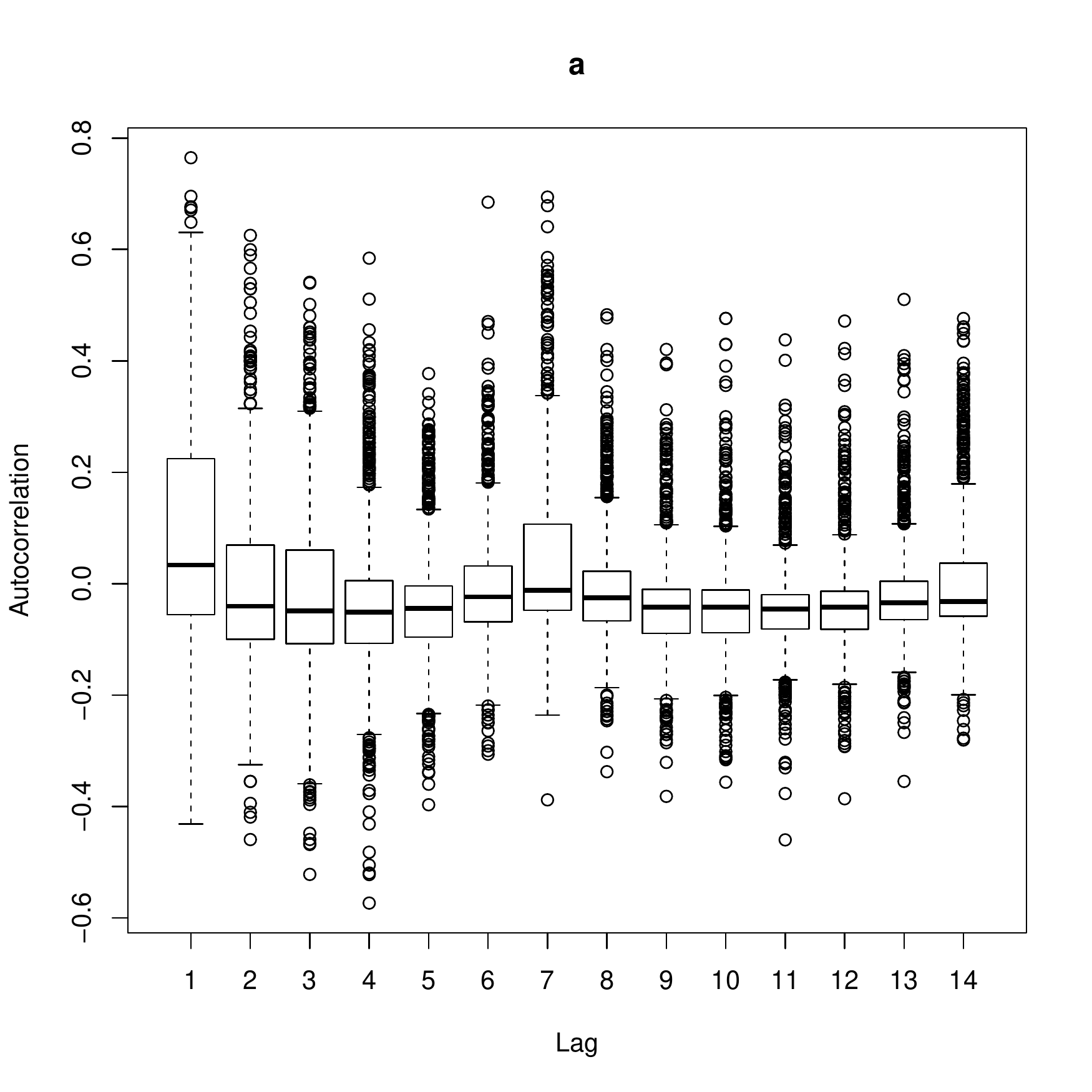}
        \refstepcounter{subfigure}\label{fig:TLFBBox}
    \end{subfigure}
    \begin{subfigure}{.45\linewidth}
        \centering
        \includegraphics[width=0.75\linewidth]{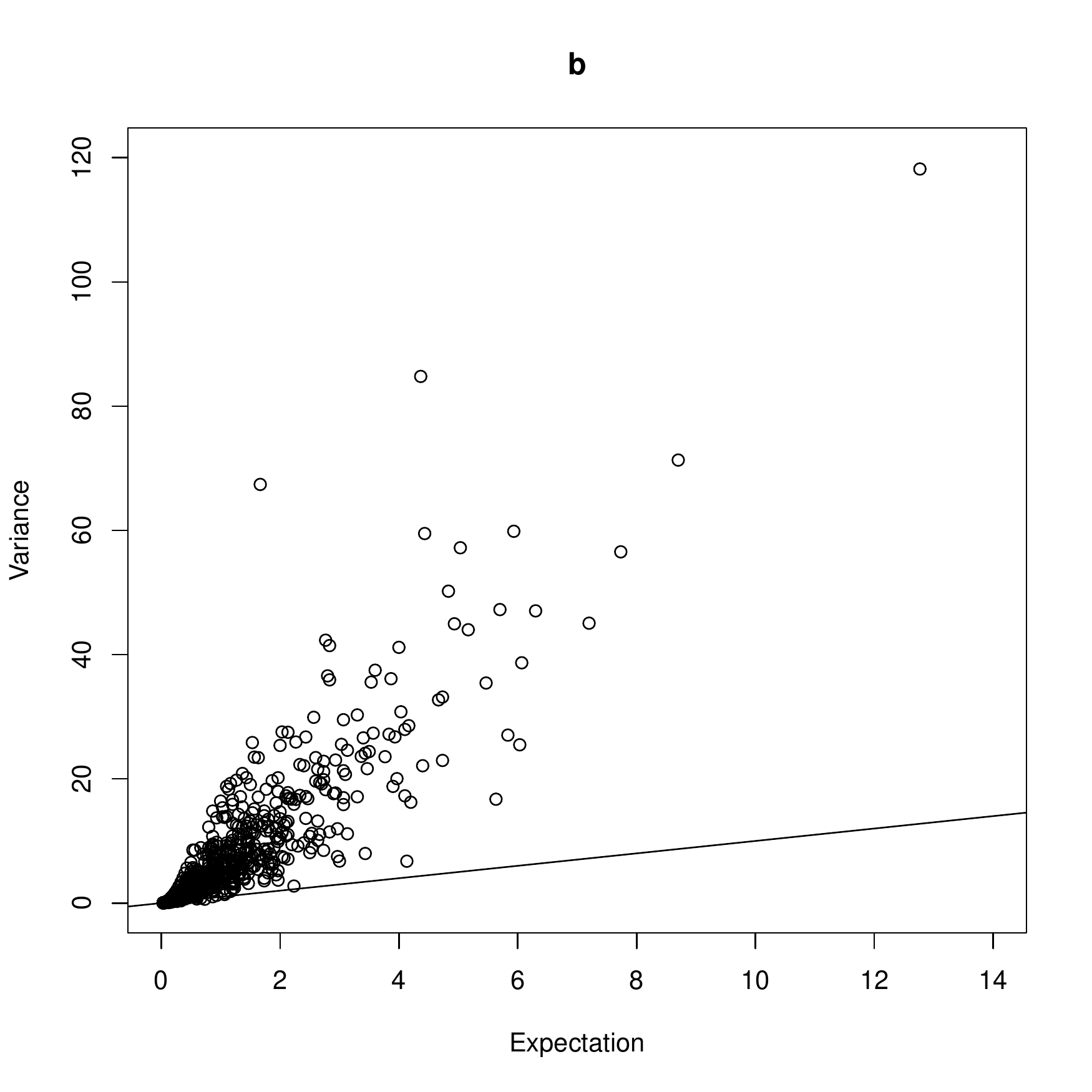}
        \refstepcounter{subfigure}\label{fig:TLFBDisp}
    \end{subfigure}\\
    \vspace{-0.5cm}
    \begin{subfigure}{.45\linewidth}
        \centering
        \includegraphics[width=0.75\linewidth]{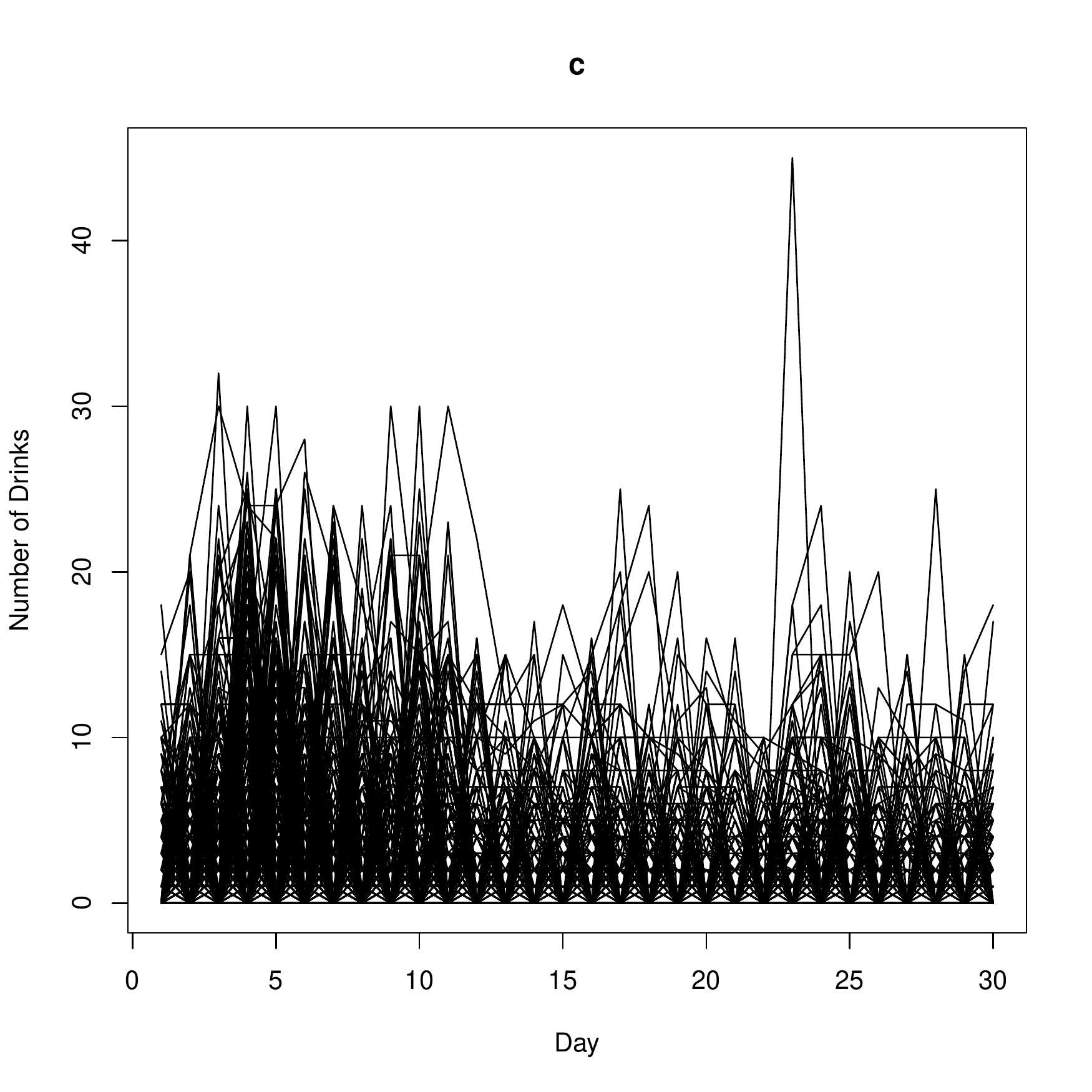}
        \refstepcounter{subfigure}\label{fig:TLFBPlot}
    \end{subfigure}
    \begin{subfigure}{.45\linewidth}
        \centering
        \includegraphics[width=0.75\linewidth]{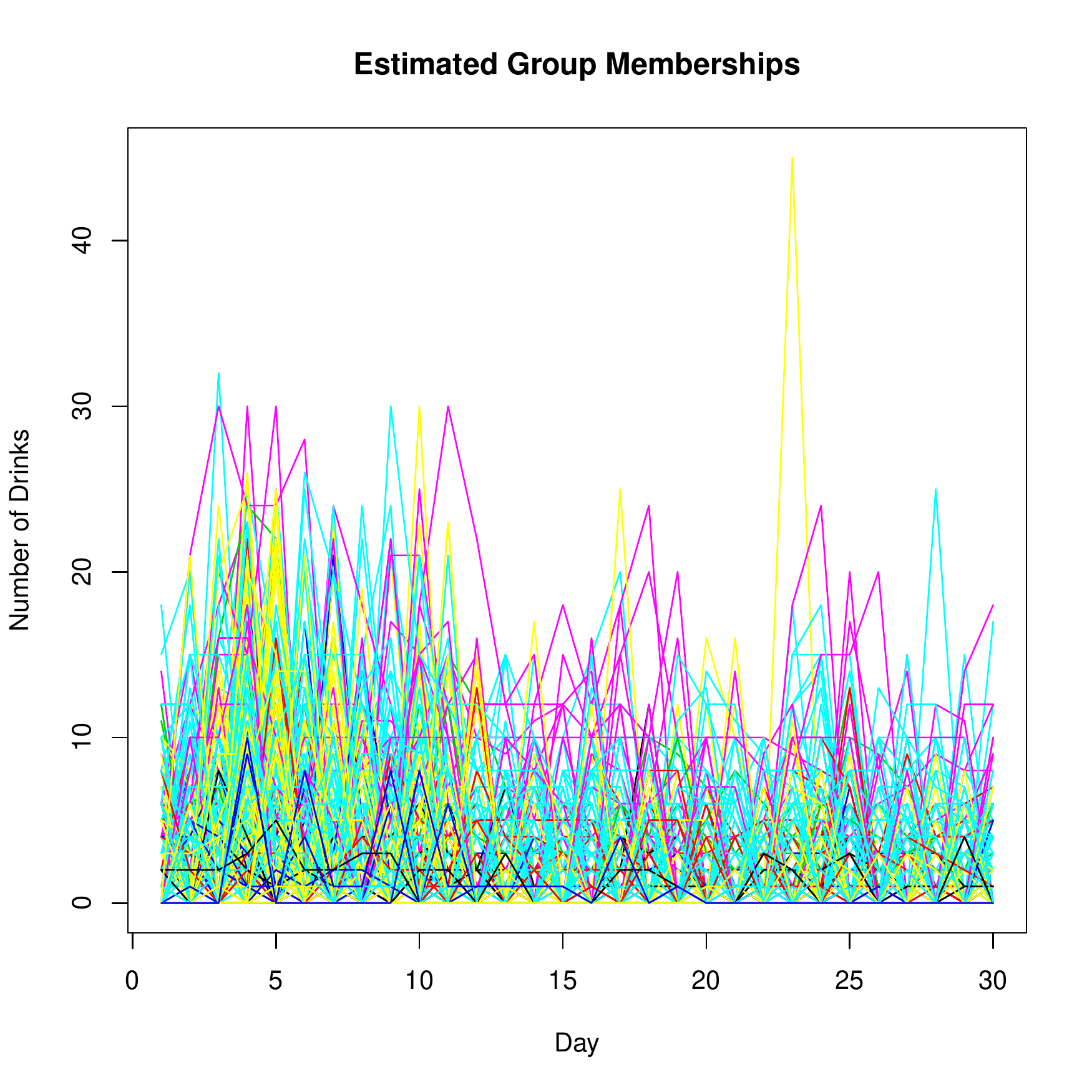}
        \refstepcounter{subfigure}\label{fig:TLFBMAP}
    \end{subfigure}\\
    \vspace{-0.5cm}
    \begin{subfigure}{\linewidth}
        \centering
        \includegraphics[width=0.3375\linewidth]{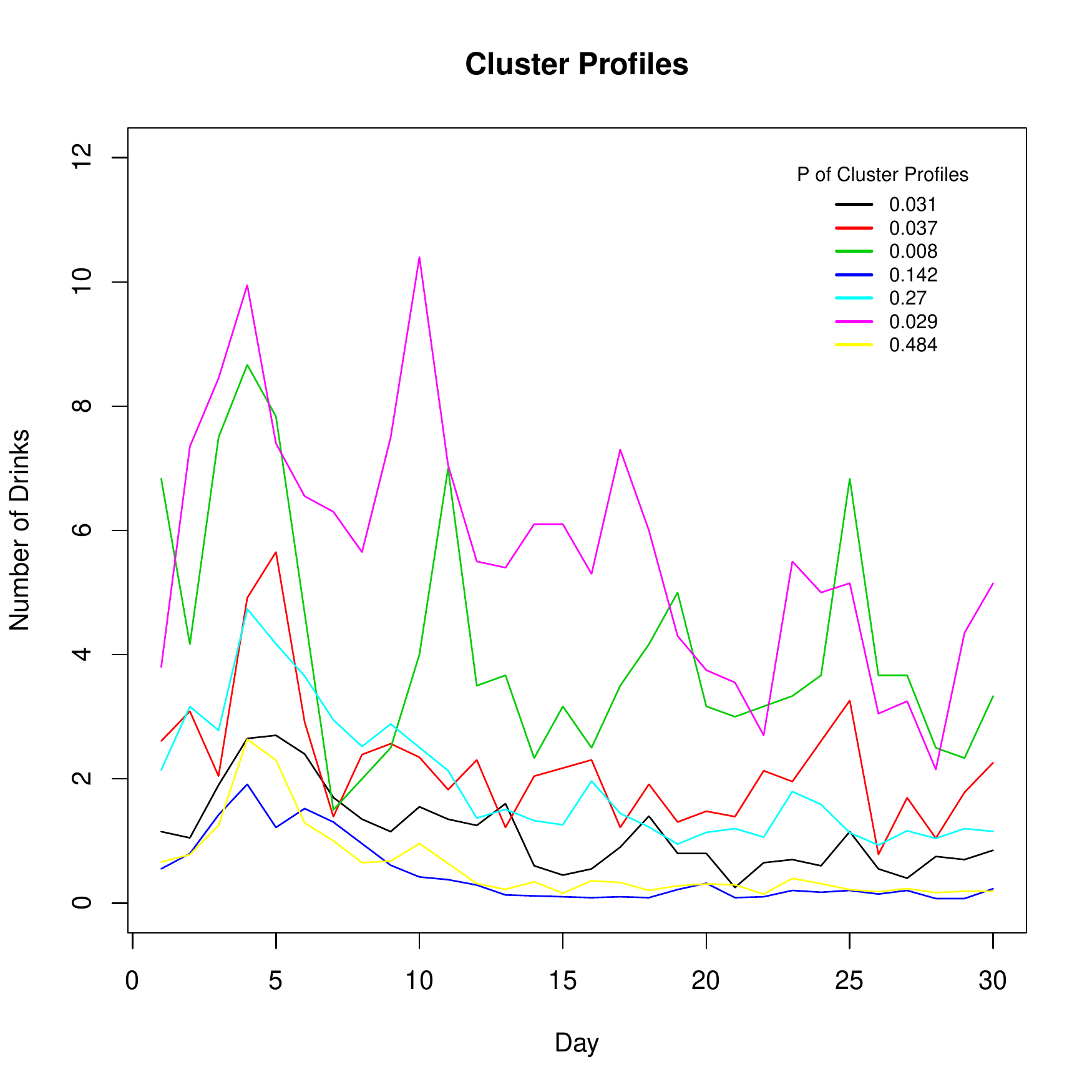}
        \refstepcounter{subfigure}\label{fig:TLFBClust}
    \end{subfigure}
    \caption{Plots of the: a) autocorrelation at multiple lag times, b) dispersion in the data, c) unknown group memberships, d) estimated group memberships, and e) cluster profiles of the estimated group memberships for the alcohol TLFB data. }
    \label{fig:TLFBResults}
\end{figure}

The data are composed of $980$ individuals  who listed their
respective number of drinks over a $30$-day period. There were
$269$ individuals who did not finish the study, and
we will only consider the $711$ individuals for which the data
were fully recorded. Taking a closer look at the data, Figure
\ref{fig:TLFBBox} shows box plots of the autocorrelations. From
these box plots,  only mixtures of INAR$(1)$ and
INAR$(7*)$ will be considered  in the model. It can
be seen from Figure \ref{fig:TLFBDisp} that overdispersion is
present in the TLFB data.  

For the alcohol TLFB data, $G=2,\ldots,8$ components are fitted using
$k$-means starting values.  The  tolerance level for
convergence is taken to be $10^{-2}$.  The BIC
selects $G=7$ components using a mixture of seven INAR$(1)$ and
zero INAR$(7*)$ processes.  This suggests that weekly
drinking habits are not very prominent in the data. 
Figure~\ref{fig:TLFBPlot} shows the raw, i.e., unlabelled, data.
Figure~\ref{fig:TLFBMAP} shows the estimated group memberships of
the TLFB data, and Figure~\ref{fig:TLFBClust} shows the respective
cluster profiles of the estimated group memberships. From the
seven cluster profiles present in Figure~\ref{fig:TLFBClust},
there seem to be individuals on very extreme ends of the
spectrum. The blue profile appears to be individuals who drank around the times of
their twenty-first birthday and returned to not drinking throughout the
remainder of the study. The yellow profile, although very similar
to the blue profile, appears to be individuals who continued
drinking lightly after the specified event. The black, red, and
turquoise profiles appear to be individuals with heavier drinking
habits, but at a variety of different quantities. The green and
purple profiles appear to be individuals with very heavy drinking
habits that occur on a regular basis. The differences in cluster
profiles could perhaps have to do with the individuals' alcohol
tolerance level, or with attendance at other social gatherings besides the time around their twenty-first birthday.

\subsubsection{Long Distance Running Strategy Data}

This long distance running strategy (LDRS) data comes from the $2012$  International Association of Ultrarunners (IAU) World Championship held in Katowice, Poland. The dataset is available at {\tt maths.ucd.ie/$\sim$brendan/data/24H.xlsx}. Ultrarunners are individuals who compete in ultra-marathons, which are considered to be any race that is longer than $26.2$ miles or $42.195$ km. The types of races include $50$km, $100$km, $6$ hour, $12$ hour, $24$ hour, and $48$ hour races. The  LDRS data considered herein  is composed of $260$ individuals who participated  in a (consecutive) $24$ hour race. In such a race, the number of cumulative laps is recorded at the end of each hour, and the individual with the greatest amount of laps at the end of the $24$ hour period is declared the winner. There were $12$ runners  who were unable to complete a single lap  and/or participate in the race.  These runners have been excluded from the data so that only the $248$ runners who completed at least one lap will be considered in this analysis.   Although variables for age, country of origin, and gender are provided, we do not consider them in our model because our goal while clustering these data is to analyze the  different running strategies  used by ultrarunners. The three main strategies to running ultra-marathons can be summarized as: running with a consistent pace for the entire race; starting with a fast pace and slowing down earlier; and starting at a consistent pace, slowing down through the middle of the race, and finishing with a fast pace.

\begin{figure}[!ht]
    \centering
    \begin{subfigure}{.45\linewidth}
        \centering
        \includegraphics[width=0.75\linewidth]{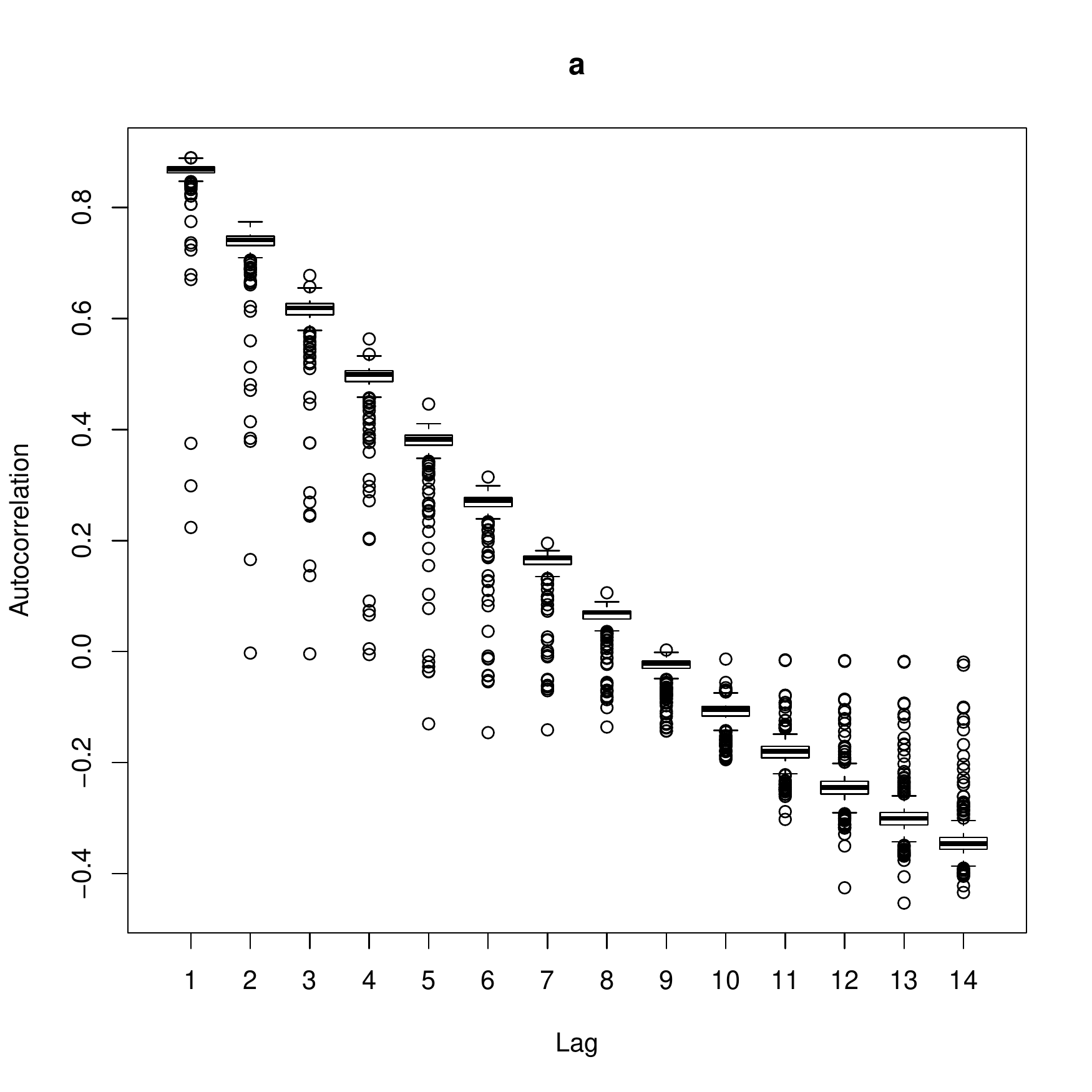}
        \refstepcounter{subfigure}\label{fig:LDBox}
    \end{subfigure}
    \begin{subfigure}{.45\linewidth}
        \centering
        \includegraphics[width=0.75\linewidth]{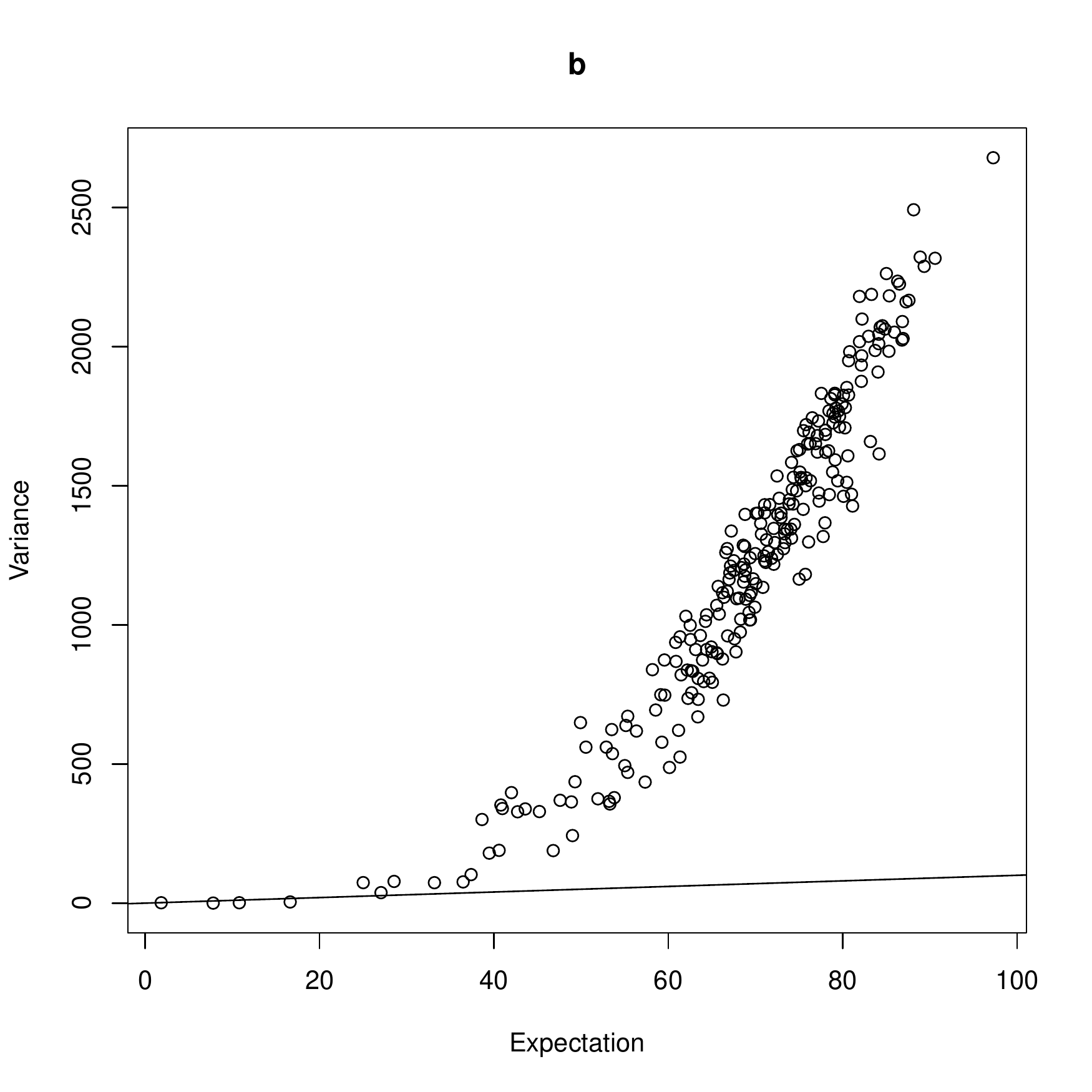}
        \refstepcounter{subfigure}\label{fig:LDDisp}
    \end{subfigure}\\
    \begin{subfigure}{.45\linewidth}
        \centering
        \includegraphics[width=0.75\linewidth]{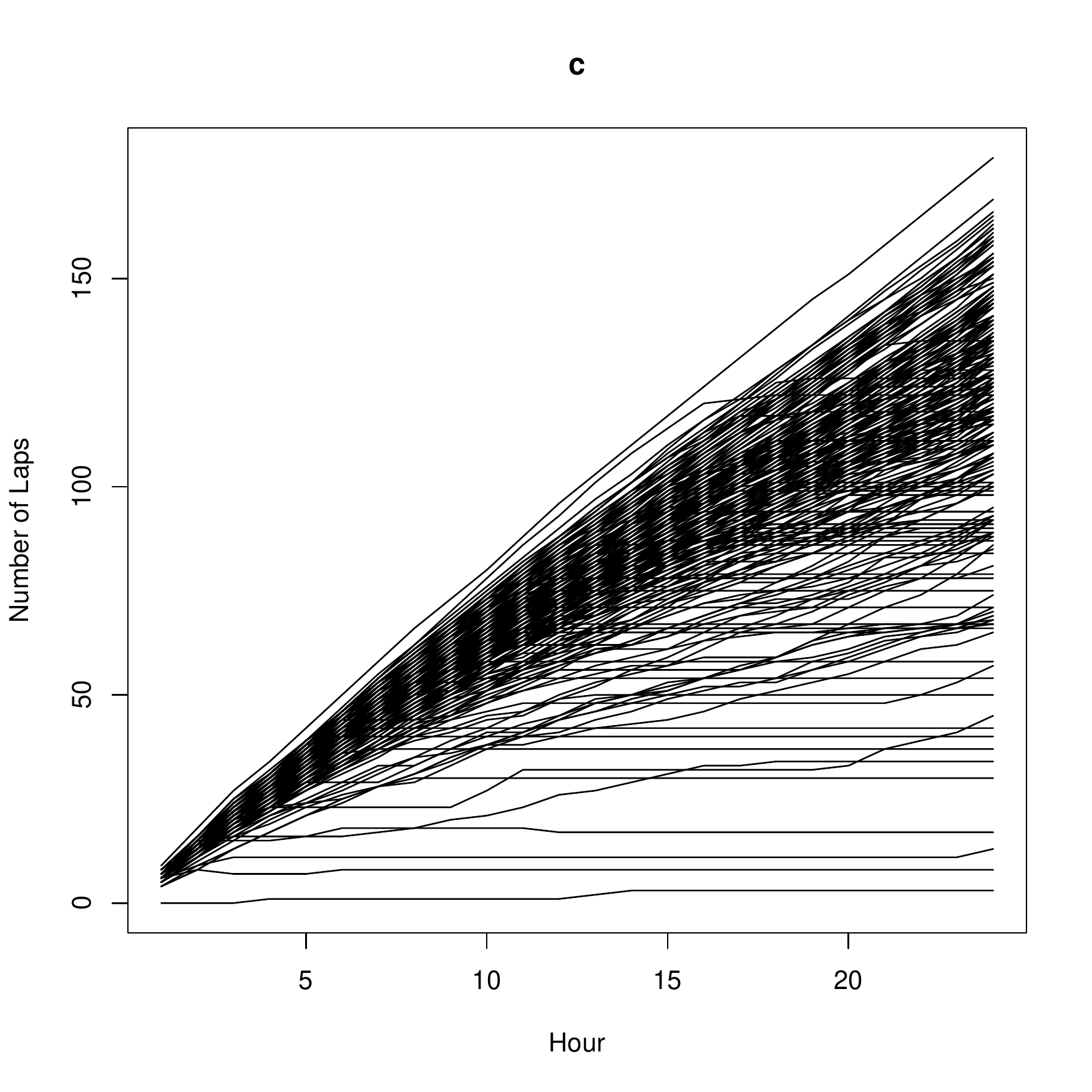}
        \refstepcounter{subfigure}\label{fig:LDPlot}
    \end{subfigure}
    \begin{subfigure}{.45\linewidth}
        \centering
        \includegraphics[width=0.75\linewidth]{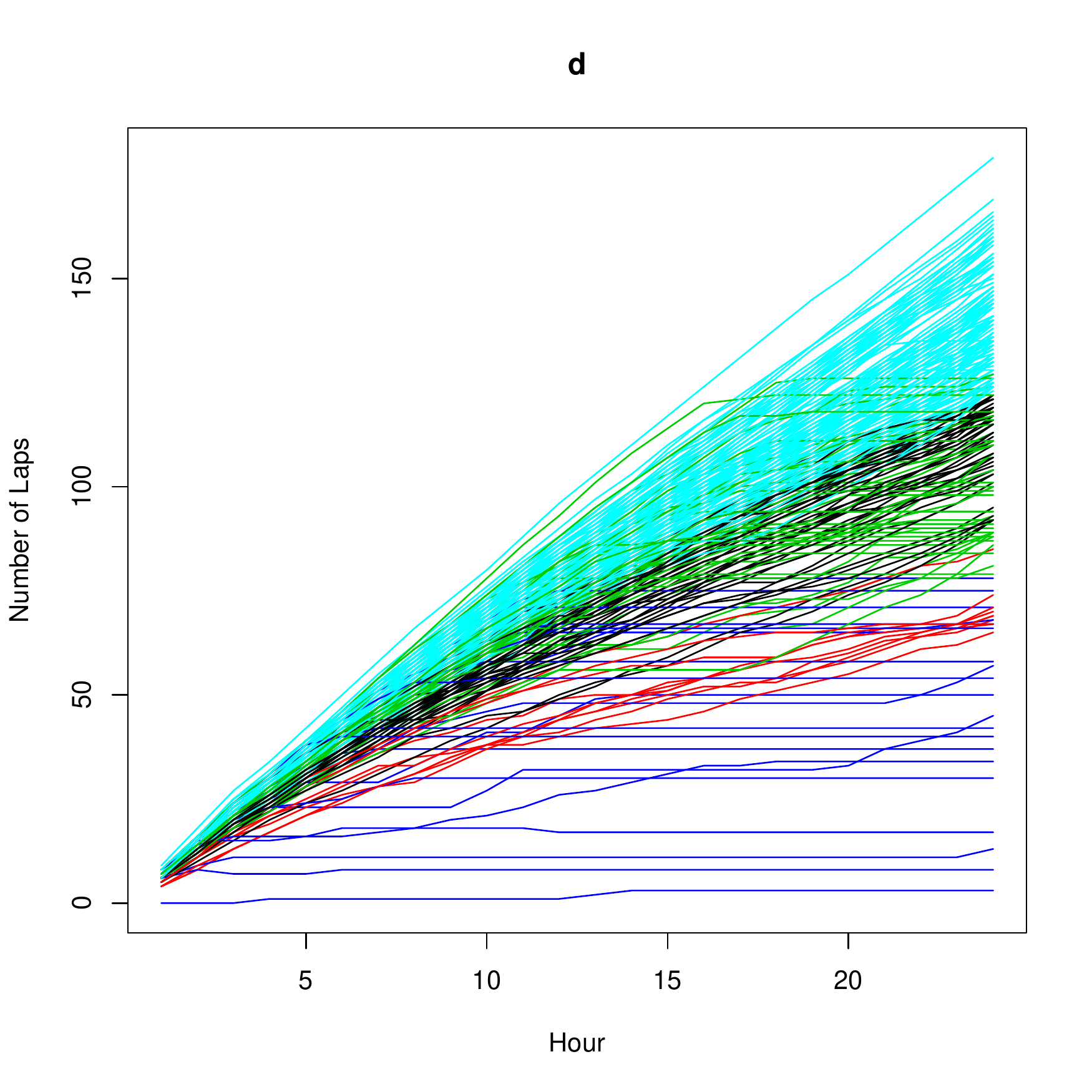}
        \refstepcounter{subfigure}\label{fig:LDMAP}
    \end{subfigure}\\
    \begin{subfigure}{\linewidth}
        \centering
        \includegraphics[width=0.3375\linewidth]{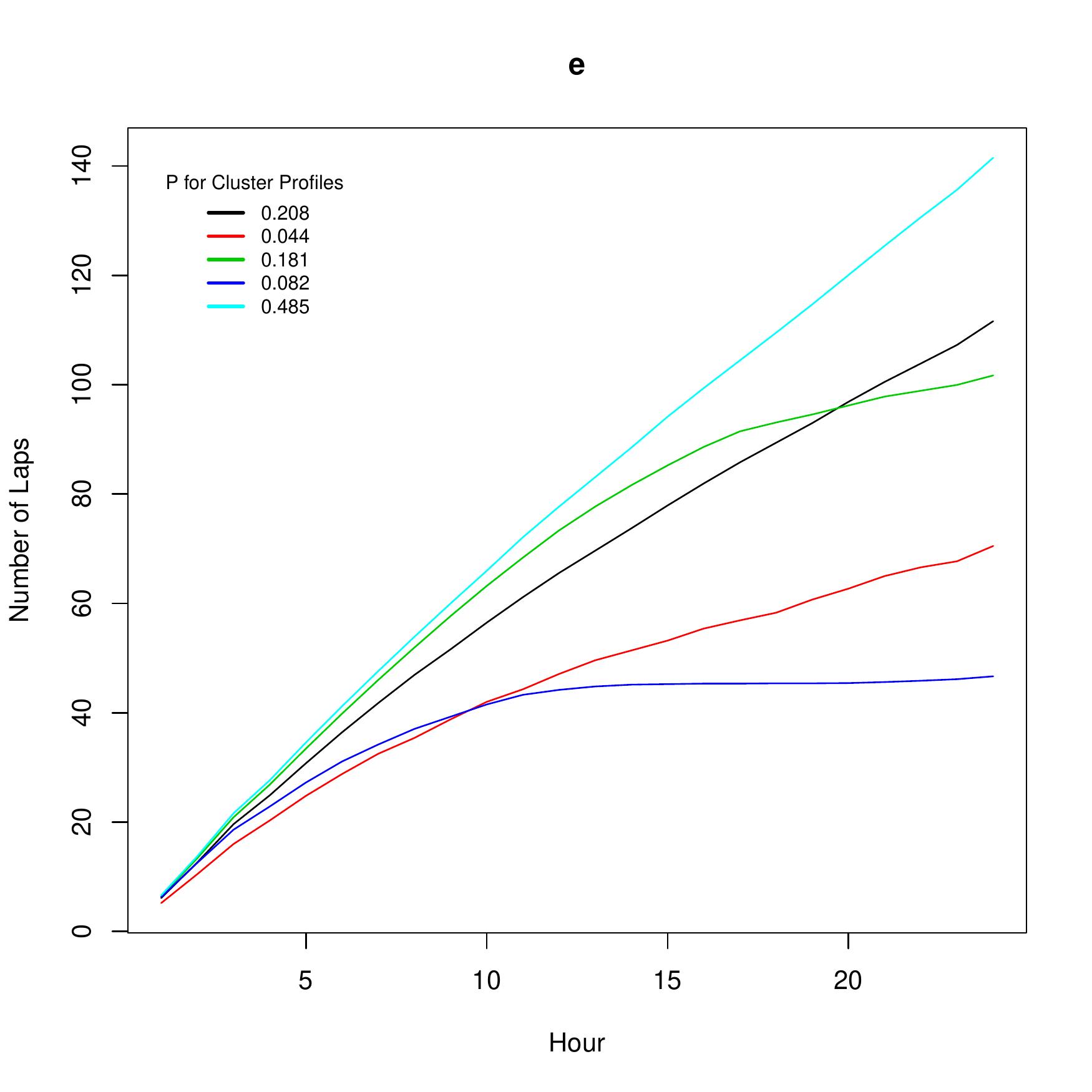}
        \refstepcounter{subfigure}\label{fig:LDClust}
    \end{subfigure}
    \caption{Plots of the: a) autocorrelation at multiple lag times, b) dispersion in the data, c) unknown group memberships, d) estimated group memberships, and e) cluster profiles of the estimated group memberships for the long distance running strategy data. }
    \label{fig:LDResults}
\end{figure}

Taking a closer look at the data, Figure \ref{fig:LDBox} shows box plots of the autocorrelations. From these box plots,  only mixtures of INAR$(1)$ and INAR$(2*)$ will be considered  in the model. It can be seen from Figure \ref{fig:LDDisp} that  overdispersion is clearly present in the LDRS data. This suggests that negative binomial distributed innovations should be considered.   

For the LDRS data, $G=2,\ldots,10$ components are fitted using
$k$-means starting values.  The  tolerance level for
convergence is again taken to be $10^{-2}$.   The BIC
selects $G=5$ components using a mixture of five INAR$(1)$ and
zero INAR$(2*)$.  This implies that runner strategies
are discovered on a lap-to-lap basis. 
Figure~\ref{fig:LDPlot} shows the raw, i.e., unlabelled, data. 
Figure~\ref{fig:LDMAP} shows the estimated group memberships of the LDRS
data and Figure \ref{fig:LDClust} shows the respective cluster
profiles of the estimated group memberships. From the five cluster
profiles present in Figure \ref{fig:LDClust}, we see that the
turquoise and black profiles are the consistent pace runners
previously mentioned. The green profile represents the runners who
start with a fast pace and slow down earlier. The red profile is
runners who start fast relative to their capabilities and slow
down throughout the middle. It looks as if these runners attempted
to pick up their pace at the end but were unsuccessful. The blue
profile looks to be runners who dropped out early due
to their running capabilities as no real strategy is present.

\section{Discussion}

A  model-based approach for clustering discrete-valued time series has been introduced. The model parameters were estimated using the EM algorithm. A stopping  criterion based on Aitken acceleration was used to determine if the model had converged. The model-based technique was applied to both simulated and real data to illustrate its clustering capabilities. Model selection was done using the BIC and, for simulated data, performance assessment was carried out using the ARI.
In the applications to simulated data, the technique performed very well for a variety of scenarios with different overlapping among the clusters. Both equidispersion and overdispersion cases were present in the simulated data. In the application to real data, two true clustering scenarios --- i.e., where there are no ``true'' groups or labels --- were considered. The technique performed appropriately and reasonable numbers of clusters were found for the obscure relationships within the data.

This novel model-based approach for clustering discrete-valued time series presents many different directions that could be taken in future work. Some of the more relevant directions to be taken include extending the INAR model to include multivariate time series of counts and using the generalized INAR$(p)$ model (see Definition~\ref{def3}) where all $p$ lags would be considered in the model. Other directions include expanding the model-based approach to include other integer-valued models, e.g., a mixture of INARCH models, and the improvement of computational aspects, e.g., the maximization step in the EM algorithm is time consuming.

While the existing literature contains other approaches for clustering time series, we have focused on the discrete nature of the data. The existing literature ignores this property and, hence, it is of interest to examine the behaviour of
distance-based and other related methods on discrete-valued data. Applying one such  method herein, we found that our approach has some advantages but a more wide ranging comparison would be useful. For example, one may attempt to use the frequency domain and apply spectral analysis to the integer valued time series for clustering purposes. 



{\small
\section*{Acknowledgements}
The authors are grateful to anonymous reviewers for their very helpful comments. This work was supported by the Canada Research Chairs program and an E.W.R.~Steacie Memorial Fellowship (McNicholas).

}


\end{document}